\documentstyle[prl,aps]{revtex}

\begin{document}
\title{Astrophysical limits on quantum gravity motivated birefringence}
\author{Reinaldo J. Gleiser and Carlos N. Kozameh}
\address{Facultad de Matem\'atica, Astronom\'\i a y F\'\i sica, 
Universidad Nacional de C\'ordoba \\
Ciudad Universitaria,  5000 C\'ordoba, Argentina}

\maketitle

\begin{abstract}
We obtain observational upper bounds on a class of quantum gravity related
birefringence effects, by  analyzing the presence of linear polarization
in the optical and ultraviolet spectrum of some distant sources. In the
notation of Gambini and Pullin we find $\chi < 10^{-3}$.
\end{abstract}




Recently two predictions from the leading theories of quantum gravity
have raised the expectations that the theories (or at least models
based on these theories) could be tested in the near future
\cite{GaPu}, \cite{A-C}. The main prediction coming from superstrings
is that the frequency dispersion relation is not linear and
consequently photons at different energies travel at different speeds.
In particular, Amelino-Camelia et al have argued that energetic photons
travel slower than soft ones \cite{A-C}. Thus, by comparing the time
of arrival of rays at different energies emitted simultaneously from
the same source, one can test the validity of this prediction. A first
step in this direction is given in \cite{biller}. A natural testing
ground to test these results are the so called GRB\'{}s
since a) they are at cosmological distances, b) some of the GRB\'{}s
last a few milliseconds and c) they emit in a continuum spectrum
(typically an E$^{-2}$ law) . So far results from the BATSE have been
unable to check this result. This is not surprising since the
difference of time of arrival for the energy range of the
BATSE\ experiment is of the order of a few microseconds, well beyond
the capability of that instrument not to mention the duration of the
burst. Two new detectors: AGILE and GLAST \cite{Agile}, to be launched
in 2002 and 2005 respectively, will be able to detect differences of
time of arrival of the order of milliseconds corresponding to an energy
difference of approximately 100 MeV. These instruments therefore will
be able to detect or at least put an upper bound on the predicted
dispersion relation.

The second prediction comes from loop quantum gravity. It was observed
by Gambini and Pullin \cite{GaPu} that if the weave states of quantum
gravity have a definite parity, then light traveling in that medium
will display birefringence,
namely, photons with left and right circular polarizations will travel
at different speeds. The predicted time of arrival difference for left
and right photons emitted simultaneously at cosmological distances is
of the same order of magnitude as for the first prediction and thus,
could be detected in the near future.

In more detail, Gambini and Pullin assume  a nonparity invariant weave, with
the result that the (vacuum) Maxwell fields satisfy equations of motion of
the form,
\begin{eqnarray}
\label{GaPu01}
\partial_t {\vec{E}} & = & - \nabla \times \vec{B} + 2 \chi \ell_P
\Delta^2 \vec{B} \nonumber \\
\partial_t {\vec{B}} & = &  \nabla \times \vec{E}  - 2 \chi \ell_P
\Delta^2 \vec{E}
\end{eqnarray}
where $\ell_P$ is the Plank length, and $\chi$ is a dimensionless
constant, that characterizes both a parity nonconservation, and a
violation of Lorentz covariance. Combining these equations, together with
$\nabla\cdot \vec{E} = \nabla\cdot \vec{B} = 0$, they obtain for $\vec{E}$
a wave propagation equation of the form,
\begin{equation}
\label{waveE}
\partial^2_t \vec{E} -\nabla^2 \vec{E} -4 \chi \ell_P \Delta^2 (\nabla
\times \vec{E})= 0
\end{equation}
and a similar equation for $\vec{B}$, where terms of higher order in $\chi
\ell_P $ have been dropped, on account of their assumed smallness. Plane
wave solutions of (\ref{waveE}), with wave vector $\vec{k}$, and given
helicity, will be of the form,
\begin{equation}
\label{waveEh}
  \vec{E}_{\pm} = \mbox{Re}( (\widehat{e}_1\pm i \widehat{e}_2)
e^{i(\Omega_{\pm} t - \vec{k} \cdot \vec{x})})
\end{equation}
with $ \widehat{e}_1 \cdot \widehat{e}_2 =0$. Consistency with (\ref{waveE}) implies
\begin{equation}
\label{biref1}
 \Omega_{\pm} = \sqrt{k^2 \mp 4 \chi \ell_P k^3} \simeq |k|(1 \mp 2 \chi
\ell_P |k|),
\end{equation}
and, $\widehat{e}_1 \cdot \vec{k} = \widehat{e}_2 \cdot \vec{k} = 0$.
Thus, the model leads to the emergence of a
birefringence effect, associated with quantum gravity corrections to the
propagation of electromagnetic waves, because the group velocity
associated with the dispersion relation (\ref{biref1}) has two branches,
one for each mode of circular polarization. In principle, this effect is
very small, corresponding roughly to a shift of one Planck length per
wavelength.

A possible way of detecting the effect, based on the difference in time
of arrival associated with the difference in group velocities was
suggested in \cite{GaPu}. In this Letter we consider a different
analysis,
which is aimed at finding upper bounds on the magnitude of the
effect. The main idea is more easily stated by restricting our attention
to an astrophysical process where photons are emitted with linear
polarization, and are detected after traveling a distance of cosmological
relevance. We place the origin of coordinates at the emission point, and
consider propagation along the $z$-direction. As a first approximation we
disregard curvature effects. Real photons cannot be represented by plane
waves, but rather by appropriate wave packets. If we assume that the
photons are emitted at times $t$ near $t=0$, with a central frequency
$\Omega_0$, and gaussian frequency width $\delta \Omega$, their wave
function may be represented by a gaussian wave packet of the form,
\begin{equation}
\label{wavefunc}
\vec{E} = \mbox{Re}\left\{{\cal{A}} e^{i \Omega_0(t-z)} \left[
  e^{-(z-v_+ t)^2 (\delta \Omega)^2} e^{-i \chi \ell_P z \Omega_0^2}
  \widehat{e}_+ \;+\;  e^{-(z-v_- t)^2 (\delta \Omega)^2} e^{ i \chi \ell_P z \Omega_0^2}
  \widehat{e}_- \right] \right\}
\end{equation}
where $\cal{A}$ is a constant,
 $ \widehat{e}_{\pm} = (\widehat{e}_1 \pm i \widehat{e}_2)/\sqrt{2}
$, and we have kept only the lowest, non trivial orders in $\chi$. The group
velocities $v_{\pm}$, corresponding to the circular polarizations $
\widehat{e}_{\pm}$, are given by,
$ v_{\pm} = 1 \mp 4 \chi \ell_P |k|
$.

Clearly, since the effect has not yet been observed, even if $v_+ \neq
v_-$, we must have $|v_+-v_-| << 1$, and $v_+ \simeq v_- \simeq 1$.
Then, the wave packets corresponding to both circular polarizations are
centered near $z= t$, and at any distance $z$ from the source, such
that,
\begin{equation}
\label{distance1}
|v_+-v_-| t \delta \Omega \simeq 8 \chi \ell_P z \Omega_0 \delta \Omega
<<1
\end{equation}
we have,
\begin{equation}
\label{rotation1}
\vec{E} \simeq \mbox{Re}\left\{{\cal{B}} e^{i \Omega_0(t-z)}  e^{-(z- t)^2
(\delta \Omega)^2} \left[
   \cos ( \chi \ell_P z \Omega_0^2 )
  \widehat{e}_1  + \sin( \chi \ell_P z \Omega_0^2)
  \widehat{e}_2 \right] \right\}
\end{equation}
Therefore, for a  sufficiently narrow (small $\delta \Omega$) packet,
at distances satisfying (\ref{distance1}), we recover the well known
rotation of the polarization plane, proportional to the distance to the
source, that characterizes optical birefringence. On the other hand, if
we consider distances $z$ such that,
\begin{equation}
\label{distance2}
 8 \chi \ell_P z \Omega_0 \delta \Omega >> 1
\end{equation}
the wave function splits into two spatially separated pieces,
corresponding to
each one of the polarization modes. This is the situation envisaged in
\cite{GaPu}, where photons corresponding to each circular polarization
would be detected with a time delay of the order of $\chi \ell_P z
\Omega_0$. But, and this is one of the main points of our discussion, when
(\ref{distance2}) holds, the photons are no longer linearly polarized. In
more precise terms, if we characterize a linear polarization detector by a
(fixed) unit vector $\vec{n}$, such that $\vec{n} \cdot \vec{k}=0$, then,
we may obtain a measure of the amount of linear polarization in the
direction of $\vec{n}$ by considering the quantity,
\begin{equation}
\label{pola1}
{\cal{P}}(\vec{n}) = < |\vec{n}\cdot \vec{E}|^2 >/< |\vec{E}|^2 >
\end{equation}
where $< >$ indicates a suitable average, for instance, we may take $< X >
= \lim_{T \rightarrow \infty}(1/T) \int_0^T X dt$. When condition
(\ref{distance1}) is satisfied, we find,
\begin{equation}
\label{pola2}
{\cal{P}}(\vec{n}) =  \cos^2 \phi = 1/2 (1+\cos (2\phi))
\end{equation}
where $ \cos \phi = \vec{n}\cdot \widehat{e}_1 $. However, when $z$ is
large enough that (\ref{distance2}) holds, we find,
 ${\cal{P}}(\vec{n}) =  1/2$, 
{\em independent} of the direction of $\vec{n}$. Therefore, if one could
be sure that the photons were emitted linearly polarized, a signal of the
presence of the birefringency effect would be the absence of this
polarization at the detector, even if no time delay measurement is
possible, at the given level of detector discrimination. In fact, this
reasoning may be further refined, because the linear
polarization is ``erased'' before a long  separation of the modes is
achieved. To see this consider a distance intermediate between
(\ref{distance1}), and (\ref{distance2}). Suppose, for example, that the
packets corresponding to right and left circular polarization essentially
superpose each other only through half of their length. Then, at the
detector, one first ``sees'' one polarization. As the packets begin their
superposition, this turns to elliptic polarization, until a linear
polarization is achieved, when the amplitudes of the packets are similar,
and from that point on the situation is reversed, the polarization becomes
again elliptic, and finally circular with the other mode. Clearly, the
observed polarization will be such that,
${\cal{P}}(\vec{n}) = 1/2+ \alpha \cos(2 \phi)$, 
with $0 \leq \alpha \leq 1/2$, where $\alpha=1/2$ corresponds to full
linear polarization and $\alpha=0$ to the absence of linear polarization,
which requires only full separation of the polarization modes.

Of course, all the previous argument relays in the knowledge of a
mechanism that certainly produces linearly polarized photons. We may,
however, turn the argument around, and ask ourselves under what conditions
it would be possible to observe a linear polarization of photons, assuming
{\em both} that they are emitted linearly polarized, {\em and} that a
birefringency effect, such as the one proposed in \cite{GaPu} takes place.
If we assume that linearly polarized photons are detected, and
unambiguously identified with a source at cosmological distance $z$,
without any significant interaction in between, we may be immediately
sure that (\ref{distance1}) is not strongly violated. Thus, if we can
measure  $z$, and $\Omega_0$, from (\ref{distance1}) we find an upper
bound on $\chi$. But, at this point, if we review the previous
derivations, we see that they refer only to essentially monochromatic
waves. Actually, photons from a given cosmological source may be observed
both in line and in continuous spectra. In the latter case, where the mechanism
at the source responsible for the linear polarization gives rise to
photons in a range of frequencies, but all polarized along the same
direction, such as synchrotron emission, or polarization by reflection
from an interstellar cloud, it is better to refer our result to the
standard definition in terms of Stokes parameters. Introducing the
quantities \cite{esto}
\begin{equation}
\label{estoques1}
{\cal{S}}_0 = < (E_x)^2> +< (E_y)^2> \;\;,\;\;
{\cal{S}}_1 = < (E_x)^2> -< (E_y)^2> \;\;,\;\;
{\cal{S}}_2 = <2 E_x E_y>
\end{equation}
the polarization is given by,
\begin{equation}
\label{estoques2}
{\cal{P}} = { ( ({\cal{S}}_1)^2 + ({\cal{S}}_2)^2)^{1/2} \over
{\cal{S}}_0}
\end{equation}

In our case, even if we assume that (\ref{distance1}) holds, (otherwise no
linear polarization would be observed), the averages must be taken over an
ensemble of wave packets of the form (\ref{rotation1}). These are of the
form, $<A> = \int {\cal{F}}(\lambda) A(\lambda) d \lambda$, 
where $\lambda = 2 \pi/k$, and ${\cal{F}}(\lambda)$ is the distribution
function for the ensemble that includes the characteristics of filters
included in the detection device. We then find, up to an overall
normalization factor,
\begin{eqnarray}
\label{pola5}
{\cal{S}}_0 & = & \int {\cal{F}}(\lambda) d \lambda \nonumber \\
{\cal{S}}_1 & = & \int {\cal{F}}(\lambda) \cos(8 \pi^2 \chi \ell_P z
/\lambda^2) d \lambda \nonumber \\
{\cal{S}}_2 & = & \int {\cal{F}}(\lambda) \sin(8 \pi^2 \chi \ell_P z
/\lambda^2)  d \lambda
\end{eqnarray}
This implies that no linear polarization may be observed if ${\cal{F}}$
does not change appreciably in a range of values of $\lambda$ such that $8
\pi \chi \ell_P z/ \lambda^2$ changes in more than several times  $\pi$. In
other words, no net linear polarization will be observed if the plane of
polarization of the different members of the ensemble (photon with
different frequencies) are rotated in angles that cover a range larger
than $\pi$. It is important to realize that what we have in mind is an
experiment where polarized photons are actually observed, quite
independently of the production mechanism. This observation puts a limit
on the possible differential rotation of the polarization planes as a
function of frequency, and, therefore, on the value of $\chi$. Since the
effect depends on $\lambda^2$, we obtain a very sensitive prove to the
presence of a birefringence of the type proposed in \cite{GaPu}. In fact,
there are available in the literature many measurements showing linear
polarization in the light from quasars, or radio galaxies, with
wavelengths in a continuous optical range.

We may take, for example, the results of Jannuzi, et.al. \cite{jannuzi} ,
that indicate a polarization larger than 10 \% in the ultraviolet for
radio galaxy 3C 256 , at a redshift $z=1.82$,  and assume, for simplicity
no change in wavelength after emission, and a flat spectrum in the region
of interest. For polarization measurements in the ultraviolet with a U
filter we take, also for simplicity, again up to an overall normalization,
\begin{equation}
\label{filter1}
{\cal{F}}(\lambda) = \exp(-(\lambda-\lambda_0)^2/(\Delta \lambda)^2)
\end{equation}
where $\lambda_0 \simeq 3500$ \AA, with $\Delta \lambda \simeq 500$ \AA.
Then, if we take $z \simeq 10^9 $ light-years, we find that the observed
polarization can be larger than  about $10 \% $, only if
\begin{equation}
\label{bound1}
\chi \leq 5 \times 10^{-4}
\end{equation}

A different type of evidence may be obtained by noticing that there are
objects at a cosmological distance that show a linearly polarized
component throughout the visible spectrum, (of the order of a few
percents) with little dependence (less than $10^o$) of the polarization
angle with wavelength \cite{brotherton}. In this case, directly from
(\ref{rotation1}), since the rotation angle is,
 $\Delta \phi_{Pol} =
4 \pi^2 \chi \ell_P z
\left(
(1 / \lambda_1^2) - (1 / \lambda_2^2) \right)$, 
if we take $\lambda_1 = 4000$ \AA, and $\lambda_2 = 8000$ \AA, $z =
10^9$ ly, and impose $\Delta \phi < 10^o$, we find,
\begin{equation}
\label{bound3}
  \chi \leq 10^{-4}
\end{equation}

It has been suggested that the presence of a birefringence of space, of
the type given by (\ref{biref1}), might be put to test in the analysis of
events where gamma rays are involved. This is because, assuming  the
characteristic parameter $\chi$ is of the order of one, such high energies
(or short wavelengths) are required in order to separate sufficiently in
time the two polarization modes, so that their separate detection becomes
technically feasible.

In the approach considered in this paper, we take (\ref{biref1}) as an
essentially phenomenological ansatz, and, instead of trying to measure
$\chi$, we analyze the possible consequences of the presence of such an
effect as regards measurements of polarization already performed. Thus,
although the lack of polarization, or the presence of a wavelength
dependence on the polarization angle of cosmological sources may be due to
many effects, the detection of significant polarization, or lack of
rotation, is possible only if the value of $\chi$ lays below a certain
upper bound. Regarding the bounds indicated in (\ref{bound1}), and
(\ref{bound3}), they should be considered as overly conservative. In fact,
in the case of (\ref{bound1}), just taking red shifts into account would
make the effective wavelength shorter leading to a smaller upper bound on
$\chi$. Notice that by taking $\lambda_0 = 1500$ \AA, the upper bound on
$chi$ is decreased by at last a factor of 10. Similarly, taking $\lambda_1
= 1500 $ \AA, the bound (\ref{bound3}) is decreased by an order of
magnitude. Further refinements on these bounds may be achieved by a more
detailed analysis of $\Delta \phi$ in polarized sources. However, the
results obtained in this paper already show that the presence of significant
polarization in the light from cosmological sources, provides important
information on the possibility of a quantum gravity birefringence effect
of the form (\ref{biref1}).  

We close this Letter with the following remarks,
\begin{itemize}
\item Equations (\ref{GaPu01}) give the
simplest coupling between quantum gravity and Maxwell theory which
includes a parity violation.
\item More important, although the results presented here are derived
from (\ref{GaPu01}), dimensional analysis of a quantum gravity induced
birefringence indicates that the change of phase of the linear
polarization vector per unit length  should be proportional to $
\ell_P/ \lambda^2$, since it should vanish in the classical limit
$\ell_P \rightarrow 0$. Thus, the result presented here should also
apply to any model that gives rise a quantum gravity induced
birefringence.
\item At present there is no reliable way to estimate the value of
$\chi$. Our results put an upper bound for this value, which might be
consistent with a more detailed calculation. Thus it would be very
important to measure polarization effects for X and gamma rays from
astrophysical sources of cosmological origin. The shorter wavelengths
of these rays would either provide a much smaller upper bound for
$\chi$, or evidence of a quantum gravity effect.
\item Since a polarization measurement is far more sensible than its time
delay counterpart, the above results indicate that evidence of quantum
gravity produced birefringence might very well be found with the
present technology.
\end{itemize}

This work was supported in part by grants of the National University of
C\'ordoba, CONICET, Fundaci\'on Antorchas, and CONICOR (Argentina). R.
J. G. is grateful to M. Heracleous for information concerning
polarized sources, and to J. Pullin, and A. Ashtekar for their valuable
comments and  hospitality at the Center for Gravitational Physics and
Geometry, at Penn State University. The authors are members of CONICET.



\begin{references}

\bibitem{GaPu}R. Gambini,  J. Pullin,
Phys. Rev. {\bf D 59}, 124021 (1999)
\bibitem{A-C} G. Amelino-Camelia, J. Ellis, N. Mavromatos, D.
Nonopoulos, and S. Sarkar, Nature (London) {\bf 393}, 763 (1998)
\bibitem{biller} S.D.Biller, et. al., Phys.Rev.Lett. {\bf 83}, 
2108-2111 (1999) 
\bibitem{Agile} See e.g., http://www.ifctr.mi.cnr.it/Agile, or  http://www..ts.infn.it/experiments/agile 
\bibitem{glast} See e.g., http://glastproject.gsfc.nasa.gov/  
\bibitem{esto} We set the Stokes coefficient ${\cal{S_3}}=0$ in
accordance with  (\ref{rotation1}).
\bibitem{jannuzi} B. T. Jannuzi, et.al. ApJ Letters {\bf 454}, L111
(1995)).
\bibitem{brotherton} Brotherton et al.  Astrophys. J. (Letters),
{\bf v.487}, p.L113 (1997)

\end{references}
\end{document}